\documentclass[aoas,MSNbibl,nameyear,seceqn,dvips]{arximspdf}
\usepackage{multirow,dcolumn}
\usepackage{subeqn,url,breakurl}
\usepackage{graphicx}

\doi{10.1214/14-AOAS721} 
\volume{8}
\issue{2}
\pubyear{2014}
\firstpage{1022}
\lastpage{1044}

\makeatletter
\newcolumntype{d}[1]{D{.}{.}{#1}}
\newcommand{\widebar}{\overline}
\newcommand{\rrvert}{\vert}
\newcommand{\llvert}{\vert}
\newtheorem{proposition}{Proposition}

\makeatother

\begin{document}
\begin{frontmatter}

\title{Statistical calibration of \lowercase{q}RT-PCR, microarray and
RNA-S\lowercase{eq} gene expression data with measurement error models\thanksref{T1}}
\runtitle{Statistical calibration of gene expression data}

\begin{aug}
\author{\fnms{Zhaonan} \snm{Sun}\ead[label=e1]{sunz@purdue.edu}},
\author{\fnms{Thomas} \snm{Kuczek}\ead[label=e2]{kuczek@purdue.edu}}
\and
\author{\fnms{Yu} \snm{Zhu}\corref{}\ead[label=e3]{yuzhu@purdue.edu}}
\runauthor{Z. Sun, T. Kuczek and Y. Zhu}
\affiliation{Purdue University}
\address{Department of Statistics\\
Purdue University\\
250N University Street\\
West Lafayette, Indiana 47907-2066\\
USA\\
\printead{e1}\\
\phantom{E-mail:\ }\printead*{e2}\\
\phantom{E-mail:\ }\printead*{e3}} 
\end{aug}
\thankstext{T1}{Supported in part by NSF Grant DMS-10-00443.}

\received{\smonth{12} \syear{2012}}
\revised{\smonth{7} \syear{2013}}

%
\begin{abstract}
The accurate quantification of gene expression levels is crucial for
transcriptome study.
Microarray platforms are commonly used for simultaneously interrogating
thousands of
genes in the past decade, and recently \mbox{RNA-Seq} has emerged as a
promising alternative. The gene
expression
measurements obtained by microarray and RNA-Seq are, however, subject
to various measurement errors.
A third platform called qRT-PCR is acknowledged to
provide more accurate quantification of gene expression levels than
microarray and RNA-Seq, but it
has limited throughput capacity. In this article, we propose to use a
system of functional
measurement error models to model gene expression measurements and
calibrate the microarray and
RNA-Seq platforms with qRT-PCR. Based on the system, a two-step
approach was developed to estimate
the biases and error variance components of the three
platforms and calculate calibrated estimates of gene expression levels.
The estimated biases and
variance components shed light on the relative strengths and weaknesses
of the three platforms
and the calibrated estimates provide a more accurate and consistent
quantification of gene
expression levels.
Theoretical and simulation studies were conducted to establish the
properties of those estimates.
The system was applied to analyze two gene expression data sets from
the Microarray Quality Control
(MAQC) and Sequencing Quality Control (SEQC) projects.
\end{abstract}

%

\end{frontmatter}

\section{Introduction}\label{secIntroduction}
The transcriptome of a cell is the entire set of mRNA molecules or
transcripts produced
by DNA transcription under certain biological or environmental conditions.
Systematic profiling of the transcriptome cannot only provide a dynamic
characterization of the cell's molecular constitution, but also shed
light on gene functional
annotation, regulatory mechanisms and transcriptional networks
underlying various biological
processes of the cell.

Since the mid-1990s, DNA microarray has served as the leading
experimental platform for
transcriptome study [\citet{Schena1995}, \citet{Lockhart1996}].
Despite its huge success, microarray is known to suffer from some
limitations such as reliance on
existing knowledge of transcript sequences, high level background noise
and limited
dynamic range of detection.

A second platform called the quantitative real-time Reverse
Transcription Polymerase Chain Reaction
(qRT-PCR) can also be used in gene expression studies.
Although well acknowledged as the most reliable gene expression measurement
technology, the throughput of qRT-PCR is low, that is, the number of
genes or transcripts that can
be measured in a single qRT-PCR experiment is limited. In addition,
qRT-PCR is also subject to
variations caused by various biological, technical and experimental
factors involved in qRT-PCR
experiments [\citet{Bustin2002}, \citet{Bustin2004}]. Currently, qRT-PCR is
used either for detecting and quantifying a small number of specific
transcript targets or as a gold
standard for validating hits or findings from other high-throughput
platforms such as microarray
[\citet{TaqManWhitePaper}].

Recently, RNA-Seq has emerged as a new experimental platform for
transcriptome profiling, and it is
believed to overcome the major limitations of microarray [\citet
{Wang2009}].
Although promising, data generated from RNA-Seq experiments still
demonstrate excessive variability [\citet{Schwartz2011}].
Therefore, RNA-Seq data need to be
normalized before used in downstream transcriptome analysis.
A number of methods have been proposed in the literature for
normalizing RNA-Seq data, including
RPKM [\citet{Mortazavi2008}], quantile-based procedures
[\citet{Bullard2010}], TMM
[\citet{Robinson2010}], mseq [\citet{Li2010}], GPseq
[\citet{Srivastava2010}], POME
[\citet{HuPome2011}] and DESeq [\citet{AndersHuber2010}].
Nonetheless, the problem of how to statistically characterize and
normalize RNA-Seq data has not
been
fully settled with satisfaction and demands further investigation
[\citet{Storey2011}].

The strengths and weaknesses of qRT-PCR, RNA-Seq and microarray can be
summarized as follows. In
terms of throughput capacity, RNA-Seq is the highest, microarray is the
second, and
qRT-PCR is the lowest, whereas, in terms of accuracy, the order from
the highest to the lowest is
qRT-PCR, RNA-Seq and
microarray. An ideal platform for transcriptome profiling should
combine the accuracy of qRT-PCR
with the
high-throughput capacity of RNA-Seq. Unfortunately, such a platform is
not currently available yet.
A natural question is whether it is possible to use statistical methods
to combine the strengths of
the three platforms and generate gene expression measurements of a
higher quality at the genome-wide
scale. The answer turns out to be positive, and the methodology that
can be applied is
\textit{statistical calibration}.

Statistical calibration is typically used for the scenario where $p$
instruments are available for
measuring
the same quantity. Among the instruments, one is more accurate but at
the same time more expensive
than the others. In order to make a less accurate instrument generate
more reliable measurement results for general use, it needs to be
calibrated by the most accurate
instrument through statistical analysis. There are two different types
of calibration, which are
\textit{absolute calibration} and \textit{comparative calibration}.
Absolute calibration assumes the most accurate instrument gives the
true value of the targeted
quantity, whereas comparative calibrate assumes the most accurate
instrument is also subject to
measurement error.
The literature on statistical calibration is
primarily focused on absolute calibration, and a variety of statistical
methods have
been developed; see \citet{Osborne1991} for a comprehensive
review. On the other hand,
comparative calibration is more challenging than absolute calibration and
is mainly discussed in the literature on measurement error models; see
\citet{Fuller1987} and
\citet{ChengNess1999}
for more thorough discussions.

The platforms for measuring gene expression levels using qRT-PCR,
microarray and RNA-Seq represent
three different instruments, with the qRT-PCR platform being the most
accurate and the most
expensive.
In this article, we propose to use a system of measurement error (ME)
models to model gene
expression measurements obtained by the qRT-PCR, microarray and RNA-Seq
platforms, and to
further use the system to calibrate
RNA-Seq and microarray measurements with qRT-PCR measurements.
A two-step approach is used to
estimate the parameters of the system of ME models, and the statistical
properties of the resulting
estimates are discussed.
Through both theoretical and simulation studies, we show that the
calibrated gene
expression measurements are more consistent and accurate than those by
any of the individual
platforms.
Furthermore, we apply the system of ME models to calibrate gene
expression data generated by
qRT-PCR, RNA-Seq
and microarray from both the Microarray Quality Control (MAQC) and
Sequencing Quality Control (SEQC)
projects (detailed information about the data is given in Section~\ref{secDatasets}), and show that
the resulting calibrated measurements provide more accurate
quantification of gene
expression levels and lead to more discoveries in gene differential
expression analysis.

In Section~\ref{secModel} we define the system of measurement
error (ME) models and describe the two-step approach
for estimating the parameters in the system. In Section~\ref{secsimulation} we present
some simulation results. In
Section~\ref{secResults}
we describe the gene \mbox{expression} data from the MAQC and SEQC projects
and discuss the results from
applying the
system of ME models to analyze the data. Section~\ref{secDiscussion} concludes the
article with further
discussion and possible future research.

\section{Measurement error models and calibration}\label{secModel}

Let $\mathcal{T}$ denote the collection of all the genes or
transcripts in the transcriptome under
study. As discussed in the \hyperref[secIntroduction]{Introduction}, RNA-Seq has the capacity to
measure the
expression levels of all genes in $\mathcal{T}$, microarray can target
thousands of genes, and
qRT-PCR generally will measure at most hundreds of genes in a single experiment.
Let $\mathcal{A}$, $\mathcal{B}$ and $\mathcal{C}$ be the
collections of genes measured by qRT-PCR,
microarray and RNA-Seq, respectively. We assume that $\mathcal
{A}\subset\mathcal{B} \subset
\mathcal{C}\subset\mathcal{T}$, and
the cardinalities of $\mathcal{A}$, $\mathcal{B}$ and $\mathcal{C}$
are $n$, $m$ and $l$,
respectively. By the assumption,
$n$ genes (i.e., those in $\mathcal{A}$) are measured by all three
platforms, $m-n$ genes (i.e.,
those in $\mathcal{B}-\mathcal{A}$) are measured by both microarray
and RNA-Seq, and $l-m$ genes
(i.e., those in $\mathcal{C}-\mathcal{B}$) are measured only by
RNA-Seq. We further assume that
commonly used platform-specific methods are used to process and
normalize raw data generated by each
platform to produce gene expression measurement data.
Some examples are the PLIER method [\citet{PLIERWhite}] for
microarray, the RPKM method for
RNA-Seq and the delta--delta $C_t$ method
for qRT-PCR. Following the convention in transcriptome studies, the
log-2 transformation is
further applied to the normalized gene expression measurements of the
three platforms, and we
refer to the resulting normalized transformed values as the qRT-PCR,
microarray and RNA-Seq gene expression
measurements, respectively, in the rest of this article.

Each type of gene expression measurement data can be generated by one
lab or multiple labs.
Based on the number of labs involved in generating one type of gene
expression measurement
data, we distinguish two scenarios, which are the single-lab scenario
and the multi-lab scenario. In
the single-lab scenario, each type of gene expression measurement data
(e.g., RNA-Seq data) is
generated by a single lab. However, the expression measurement data of
different types are not
necessarily generated from the same lab.
In the multi-lab scenario at least one type of gene expression data is
generated by more than one
lab.
Due to limited space, we treat only the single-lab scenario in this
section, and the
multi-lab scenario will be discussed in Section~\textup{S.7} of the supplementary material [\citet{SunKuczekZhu2014s1}].

\subsection{Single-lab scenario}
Suppose that the genes in $\mathcal{A}$, $\mathcal{B}-\mathcal{A}$
and $\mathcal{C}-\mathcal{B}$
are labeled by integers from $1$ to $n$, $n+1$ to $m$, and $m+1$ to
$l$, respectively. We assume
each platform generates one expression measurement for each gene. When
technical replicates are
available, their average value is used as the expression measurement of
a gene.
For $1\le j\le n$, let $X_{j}$ denote the expression measurement of
gene $j$ by qRT-PCR.
Similarly, let $Y_{j}$ denote the expression measurement of gene $j$ by
microarray for $1\le
j \le m$, and $Z_{j}$ the expression measurement of gene $j$ by RNA-Seq
for $1\le j \le
l$. Note that $X_j$, $Y_j$ and $Z_j$ are all in log-2 scales.

As discussed in the \hyperref[secIntroduction]{Introduction}, gene expression measurements produced
by the three platforms are
all subject to measurement errors.
We propose to use the following measurement error models to
characterize $X_{j}$, $Y_{j}$ and
$Z_{j}$, respectively,
\begin{subequations}
\begin{eqnarray}
\label{eqmodel-single-pcr} X_{j} & =&\mu_j + \varepsilon_{1j}\qquad (1\le j \le n ),
\\
\label{eqmodel-single-microarray} Y_{j} & =& \alpha_2 + \beta_2
\mu_{j} + \varepsilon_{2j}\qquad (1\le j \le m ),
\\
\label{eqmodel-single-rnaseq} Z_{j} & =& \alpha_3 + \beta_3
\mu_{j} + \varepsilon_{3j}\qquad (1 \le j \le l ). \label{eqmodel-single-lab}
\end{eqnarray}
\end{subequations}
The above models are not simple linear models because none of the terms
on the right-hand sides of
the equations is observed, and terms $\beta_2 \mu_j$ and $\beta_3
\mu_j$ are cross
terms of unknown quantities. Note that $n < m < l$, that is, the three
platforms measure different
subsets of genes. Next, we will
discuss the terms in these three models in detail and then propose the
estimation method.

In model (\ref{eqmodel-single-pcr}), $\mu_j$ is the true expression
level of gene $j$ and
$\varepsilon_{1j}$ is the random error due to the qRT-PCR platform.
Depending on the normalization method, \mbox{qRT-PCR} can lead to either
absolute quantitation or
relative quantitation of a gene's expression level [\citet
{Pfaffl2004}, Chapter~3].
Therefore, the interpretation of~$\mu_j$ depends on whether absolute
or relative quantitation is
used in an experiment. For
simplicity, in this article, we do not further distinguish these two
quantitation methods and simply
refer to $\mu_j$ as the true expression value of gene $j$. We assume
that for $1\le j\le n$,
$\varepsilon_{1j}$'s are i.i.d. $N(0, \sigma_1^2)$.
The variance of $X_{j}$ is $\sigma_1^2$, representing the
reproducibility of the qRT-PCR platform.
According to the model, the qRT-PCR measurement $X_{j}$ is unbiased
with respect to $\mu_j$.

In model (\ref{eqmodel-single-microarray}), $\varepsilon_{2j}$ is the random
error due to the microarray platform. For $1\le j\le m$, we assume
$\varepsilon_{2j}$'s are i.i.d.
$N(0, \sigma_2^2)$. The other two terms gives the mean measurement,
that is,
$E(Y_{j})=\alpha_2 +\beta_2 \mu_j$.
In other words, the expectation of the microarray measurement of gene
$j$ is assumed to be a linear
function\vspace*{2pt} of~$\mu_j$, where $\alpha_2$ and $\beta_2$ are the
intercept and slope, respectively. The
variance of $Y_{j}$ is~$\sigma_2^2$. Note
that due to the differences in technologies and platform specific
normalization methods, gene
expression measurements from different platforms are in different
scales and may have
shifts. Therefore, $\alpha_2$ and $\beta_2$ represent the shift and
scale of the microarray
measurements relative to the qRT-PCR measurements.
When comparing the reproducibilities of the qRT-PCR and microarray
platforms, $Y_j$ needs to be
transformed to $\widetilde{Y}_j= (Y_j-\alpha_2)/\beta_2$, which is of
the same scale as $X_j$.
The variance of $\widetilde{Y}_j$, which is $\sigma_2^2/\beta_2^2$, is
referred to as the
reproducibility of the microarray platform.

In model (\ref{eqmodel-single-rnaseq}), the mean of the RNA-Seq measurement
is also assumed to be a linear function of $\mu_j$, that is,
$E(Z_{j})=\alpha_3+\beta_3 \mu_j$, and
$\varepsilon_{3j}$ is the random error due to the RNA-Seq platform.
We assume that for $1\le j\le l$, $\varepsilon_{3j}$'s are i.i.d. $N(0,
\sigma_3^2)$. The
intercept $\alpha_3$ and the slope $\beta_3$ represent the shift and\vspace*{1pt}
scale of \mbox{RNA-Seq} measurements
relative to the qRT-PCR measurements. The variance of $Z_{j}$ is~$\sigma_3^2$.
Similar to the microarray platform, we refer to $\sigma_3^2/\beta
_3^2$ as the reproducibility
of the \mbox{RNA-Seq} platform.

Different measurement platforms generally have different dynamic ranges
of detection.
This is also the case for the qRT-PCR, microarray and RNA-Seq
platforms. The models proposed above
may hold for only a range that fits all three platforms.
Therefore, when applying the models in practice, a proper range of expression
levels needs to be used, and genes with extremely low or high
expression levels need to be
excluded. Random measurement errors are typically heteroscedastic, that is,
the variance of a random measurement error depends on the magnitude of
the targeted quantity.
Notice
that the errors in the models above are assumed to homoscedastic. The
justification for assuming
homoscedastic errors is twofold. First, the gene expression
measurements $X_j$, $Y_j$ and $Z_j$
are the log-2
values of the original measurements produced by the three platforms,
respectively, and the log-2
transformation is known to mitigate heteroscedasticity. Second, as
previously discussed, the
proposed models will be applied to genes with expression levels within
a certain proper range, which
further alleviates the concern of heteroscedasticity. Nonetheless, when
applying the models,
diagnostic analysis always needs to be performed.

When $\mu_j$'s in models (\ref{eqmodel-single-pcr})--(\ref{eqmodel-single-rnaseq}) are assumed
to be unknown fixed quantities, the measurement error models are said
to be \textit{functional}; and
when $\mu_j$'s are assumed to be \textit{i.i.d.} random variables, the
models are said to be \textit{structural}. In this article, $\mu_j$'s are gene expression values and
considered fixed. Therefore,
the models defined above are functional.
Following the convention in the literature, we call the $\mu_j$'s \textit{incidental parameters} $k$
and the other parameters ($\alpha_i$, $\beta_i$ for $2\le i \le3$,
and $\sigma_i^2$
for $1\le i \le3$) \textit{structural parameters}.

In general, when $p$ measurement platforms or instruments are used to
measure the same quantity,
it requires $p$ measurement error models to characterize the
measurement results. For ease of
discussion, the $p$ measurement error models are said to form a system
of ME models of order $p$.
When the measurement error models are structural, the system is said to
be structural; and when the
models are functional, the system is said to be functional. Hence, models
(\ref{eqmodel-single-pcr})--(\ref{eqmodel-single-rnaseq}) defined
above form a functional system of order 3, and any two of them form a
functional system of order 2.

\subsection{Parameter estimation}\label{secsingle-estimation}
The statistical literature on measurement error models is focused
primarily on
systems of order 2. \citet{Reiersol1950} showed that structural\vadjust{\goodbreak}
systems of order 2 are not
identifiable under the normality assumption,
unless additional information such as the ratio between the variances
of the two measurement
instruments is available.
For structural systems of order higher than~2 (e.g., $p=3$),
\citet{Barnett1969} showed they become
identifiable with no further information needed, and the maximum
likelihood estimates (MLEs) of the
structural parameters can be obtained.\looseness=1

Parameter estimation for functional systems of ME models is more
difficult than that for structural
systems due to the presence of the incidental parameters, $\mu_{j}$'s.
First, under the normality
assumption on the involved measurement errors, the likelihood function
of a functional system
becomes unbounded [\citet{KendallStuart1979}].
\citet{Solari1969} showed that the likelihood function of a
functional system of order 2
does not have a local maximum.
We have found that this is also generally true for functional systems
of order 3. Therefore, the
maximum likelihood method fails to produce proper estimates for both
structural and incidental
parameters of a functional system of ME models.

The relationship between structural and functional systems was
discussed in the early literature on
measurement error models. Much attention has been given to the connection
between the identifiability of structural systems and the existence of
proper estimates of the
structural parameters in functional systems. \citet{Gleser1983}
showed that when a structural
system is identifiable, consistent estimates of the structural
parameters in the corresponding
functional model exist. Further, he suggested that in most cases, a
good way to find a consistent
estimator for the functional model is to directly verify whether the
consistent estimator in the
structural model is also consistent in the corresponding functional model.

Based on the discussion above, a two-step approach can be used to
estimate the parameters of the
functional system of order 3 defined above for the \mbox{qRT-PCR}, microarray
and RNA-Seq measurements.
In the first step, the genes with measurements from all three platforms
(i.e., genes in
$\mathcal{A}$) are used to obtain the MLEs of the structural
parameters under the
assumption that these $\mu_j$'s are i.i.d. $N(\mu, \sigma^2)$.
In the second step, the estimates obtained in the first step are used in
place of their corresponding structural parameters in the functional ME
system, and then the
generalized least squares method is used to obtain estimates of the
incidental parameters (or gene
expression levels).

\subsubsection*{Structural parameters}
We use the expression measurements of the genes in $\mathcal{A}$ to estimate
the structural parameters. Let $\widebar{X}_{.}$, $\widebar{Y}_{.}$ and
$\widebar
{Z}_{.}$ be the averages
of ${X}_{j}$, ${Y}_{j}$ and ${Z}_{j}$ over all genes in $\mathcal{A}$.
The sample variances and covariances for $\{{X}_{j}\}_{1\le j \le n}$,
$\{{Y}_{j}\}_{1\le j \le n}$ and $\{{Z}_{j}\}_{1\le j \le n}$ are
denoted as $S_{xx},
S_{yy}, S_{zz}, S_{xy}, S_{xz}$ and $S_{yz}$, respectively.
Following the first step of the two-step approach discussed previously,
the estimates of the
structural parameters are given as follows:
\begin{subequations}
\begin{eqnarray}
\label{eqestimate-structural}
\qquad\quad \hat{\beta}_2 &=&\frac{S_{yz}}{S_{xz}},\qquad
\hat{\beta}_3 = \frac{S_{yz}}{S_{xy}},\qquad
\hat{\alpha}_2 = \widebar{Y}_{.} - \hat{\beta}_2\widebar{X}_{.},
\nonumber\\[-8pt]\\[-8pt]
\hat{\alpha}_3 &=& \widebar{Z}_{.} -\hat{\beta}_2 \widebar{X}_{.},\nonumber
\\
\hat{\sigma}_1^2 &=& \biggl(S_{xx} -\frac{S_{xy}S_{xz}}{S_{yz}} \biggr),\qquad
\hat{\sigma}_2^2 =\biggl(S_{yy} - \frac{S_{xy}S_{yz}}{S_{xz}} \biggr),
\nonumber\\[-8pt]\\[-8pt]
\hat{\sigma}_3^2 &=& \biggl(S_{zz} - \frac{S_{yz}S_{xz}}{S_{xy}} \biggr).\nonumber
\end{eqnarray}
\end{subequations}
Note that when deriving the above estimates, the nonnegativity
constraints on the variance
estimates were not enforced due to two considerations. First, we can
enforce the constraints by
using more complicated estimates such as the ones given in \citet
{CarterFuller1980}. However, in
practice, the constraints are usually satisfied automatically and,
thus, the benefit of using the
more complicated estimates is minimal. Second, gene expression data are
often noisy, and the true
values of the variance components are unlikely to be close to 0. The
violation of the
constraints by the variance components estimates indicates either the
estimates are not reliable due
to insufficient sample size or the model assumptions are invalid and
need to be reconsidered.
The enforcement of the constraints may miss the opportunity of
identifying these potential pitfalls.

For convenience, let $\bolds{\theta} = (\alpha_2, \alpha_3,
\beta_2, \beta_3, \sigma_1^2, \sigma_2^2, \sigma_3^2 )^{\top
}$ be the vector of the structural parameters
and $\hat{\bolds{\theta}}$ the estimate of $\bolds{\theta}$. The
following proposition
establishes the asymptotic distribution of $\hat{\bolds{\theta}}$
under the
functional system of ME models as $n$ goes to $\infty$.

%
\begin{proposition} \label{teoproposition1}
Assume the functional system of ME models is true and the following
limits exist,
\begin{equation}
\label{eqlimiting-assumption} \bar{\mu} = \lim_{n \rightarrow\infty}
\frac{1}{n} \sum
_{j=1}^n \mu_{j}\quad\mbox{and}\quad\Delta= \lim_{n \rightarrow\infty} \frac{1}{n} \sum
_{j=1}^n (\mu_{j} - \bar{\mu}
)^2 > 0.
\end{equation}
Then, as $n$ goes to $\infty$,
$ \sqrt{n} (\hat{\bolds{\theta}} - \bolds{\theta} )
\stackrel{D}{\longrightarrow} N (\mathbf{0}, \Gamma_{\bolds
{\theta}} )$,
where $\stackrel{D}{\longrightarrow}$ means converge in distribution
and $\Gamma_{\bolds{\theta}}$ is
the variance--covariance matrix with its
explicit expression given in the supplementary material [\citet
{SunKuczekZhu2014s1}], Section~\textup{S.1.2.}
\end{proposition}

The proof of Proposition \ref{teoproposition1} is outlined in Section~\textup{S.1.1} of the supplementary material [\citet{SunKuczekZhu2014s1}]. The assumptions on the
mean and variance of $\mu_j$'s in
Proposition~\ref{teoproposition1} appear to be
reasonable, because $\mu_j$'s are the true expression levels of genes
in a given cell line or tissue
sample.
Note that $\Delta$ appears in the denominator of the variances of
$\hat\alpha_2$, $\hat\alpha_3$,
$\hat{\beta}_2$ and $\hat{\beta}_3$. Therefore, the estimates of the
structural parameters will attain
higher accuracy when $\mu_j$'s in $\mathcal{A}$ are more spread out.
The asymptotic property of $\hat{\bolds{\theta}}$ is stated as $n$
goes to $\infty$. In practice, this property will hold approximately
when $n$ is sufficiently
large. In our simulation study, we found that when $n \ge150$, the
Mean Square Errors (MSE) of
$\hat{\bolds{\theta}}$ are close to zero (see Figure S1 and Figure
S2 in the supplementary material [\citet{SunKuczekZhu2014s1}], Section~\textup{S.4}) and the
standard errors of
$\hat{\bolds{\theta}}$ are within 10\% of the true
parameter values.

\subsubsection*{Incidental parameters}
How to best estimate the incidental parameter $\mu_j$ depends on whether
gene $j$ is in $\mathcal{A}$, $\mathcal{B} - \mathcal{A}$ or
$\mathcal{C}-\mathcal{B}$.
We consider these three cases separately. For each case, we
follow the second step of the two-step approach discussed previously.
First, we replace the
structural parameters by their estimates, and then we apply the
generalized least squares
method to obtain the estimate of~$\mu_j$.

Depending on whether gene $j$ is in $\mathcal{A}$, $\mathcal{B} -
\mathcal{A}$ or
$\mathcal{C}-\mathcal{B}$, given the structural parameters, the
measurement error models involving
$\mu_j$ can be rewritten as a linear model of $\mu_j$ (see the
supplementary material
[\citet{SunKuczekZhu2014s1}], Section~\textup{S.2}).
Replacing the structural parameters $\bolds{\theta}$ by their
estimates $\hat{\bolds{\theta}}$ and
applying the generalized least squares method, the estimates of $\mu
_j$ for $j$ in $\mathcal{A}$,
$\mathcal{B} - \mathcal{A}$ and $\mathcal{C}-\mathcal{B}$ are given
as follows:
\begin{subequations}
\begin{eqnarray}
\label{eqscm-prediction-xyz} \qquad\hat{\mu}_{j}^{xyz}
& =& \frac{{X}_{j}/\hat{\sigma}_1^2 + \hat{\beta}_2 ({Y}_{j} - \hat
{\alpha}_2 )/\hat{\sigma}_2^2
+ \hat{\beta}_3 ({Z}_{j} - \hat{\alpha}_3 )/ \hat{\sigma}_3^2
}{1/\hat{\sigma}_1^2 + \hat{\beta}_2^2/\hat{\sigma}_2^2 + \hat{\beta
}_3^2/\hat{\sigma}_3^2}\qquad\mbox{for }j \in\mathcal{A};
\\
\label{eqscm-prediction-yz} \hat{\mu}_{j}^{yz} & =& \frac{ \hat{\beta}_2
({Y}_{j} - \hat{\alpha}_2 )/\hat
{\sigma}_2^2 + \hat{\beta}_3
({Z}_{j} - \hat{\alpha}_3 )/\hat{\sigma}_3^2 }{ \hat{\beta}_2^2/\hat
{\sigma}_2^2 +
\hat{\beta}_3^2/\hat{\sigma}_3^2}\qquad\mbox{for }j \in\mathcal{B - A};
\\
\label{eqscm-prediction-z} \hat{\mu}_j^{z} & =& ({Z}_{j} -
\hat{\alpha}_3 )/\hat{\beta}_3\qquad\mbox{for }j \in \mathcal{C - B}.
\end{eqnarray}
\end{subequations}
We call $\hat{\mu}_{j}^{xyz}$, $\hat{\mu}_{j}^{yz}$ and $\hat{\mu
}_{j}^{z}$ the calibrated
estimates of $\mu_{j}$ for $j$ in $\mathcal{A}$, $\mathcal{B} -
\mathcal{A}$ and
$\mathcal{C}-\mathcal{B}$, respectively. The quality of the
calibrated estimates of $\mu_j$
depends on how accurate the estimates of the structural parameters are, which
further depends on~$n$, the number of genes in $\mathcal{A}$. The
properties of $\hat{\mu}_{j}^{xyz}$,
$\hat{\mu}_{j}^{yz}$ and $\hat{\mu}_{j}^{z}$ as $n$ goes to $\infty
$ are given in the following
proposition.

%
\begin{proposition} \label{teoproposition2}
Assume the functional system of the ME models is true and the limits in
(\ref{eqlimiting-assumption})
exist.
\begin{longlist}[(ii)]
\item[(i)] As\vspace*{1pt} $n$ goes to $\infty$, $\hat{\mu}_j^{xyz}$
asymptotically follows the normal
distribution $N (\mu_j, \gamma_{\mathcal{A}} )$ for $j
\in\mathcal{A}$, $\hat{\mu}_j^{yz}$
asymptotically follows the normal distribution $N (\mu_j, \gamma
_{\mathcal{B-A}} )$ for $j \in
\mathcal{B-A}$, and $\hat{\mu}_j^{z}$ asymptotically follows the normal
distribution $N (\mu_j, \gamma_{\mathcal{C-B}} )$ for $j
\in\mathcal{C-B}$, where
$\gamma_{\mathcal{A}}= (1/\sigma_1^2 + \beta_2^2/\sigma_2^2
+ \beta_3^2/\sigma_3^2 )^{-1}$, $\gamma_{\mathcal{B} -
\mathcal{A}} = ({\beta}_2^2/{\sigma^2_2} + {\beta}_3^2/{\sigma
^2_3} )^{-1}$ and $\gamma_{\mathcal{C}-\mathcal{B}} =
(\beta_3^2/\sigma^2_3 )^{-1}$.
\item[(ii)] The variances of $\hat{\mu}_j^{xyz}$, $\hat{\mu}_{j}^{yz}$ and
$\hat{\mu}_{j}^{z}$ admit the following first order expansions:
\begin{subequations}
\begin{eqnarray}
\label{eqscm-prediction-xyz-var} \operatorname{Var}\bigl(\hat{\mu}_j^{xyz}\bigr
) & \approx& \gamma_{\mathcal{A}} +
n^{-1} \omega_{\mathcal{A}} (\bolds{\theta}, \mu_j )\qquad\mbox{for }j \in\mathcal{A},
\\
\label{eqscm-prediction-yz-var} \operatorname{Var} \bigl(\hat{\mu
}_{j}^{(yz)}\bigr) &
\approx&\gamma_{\mathcal{B} -
\mathcal{A}} + n^{-1} \omega_{\mathcal{B} - \mathcal{A}} (\bolds{
\theta}, \mu_j )\qquad\mbox{for }j \in\mathcal{B - A},
\\
\label{eqscm-prediction-z-var} \operatorname{Var}\bigl(\hat{\mu}^{z}_j\bigr) &
\approx&\gamma_{\mathcal{C}-\mathcal{B}} + n^{-1} \omega_{\mathcal{C} -
\mathcal{B}} (\bolds{
\theta}, \mu_j )\qquad\mbox{for }j \in\mathcal{C - B},
\end{eqnarray}
\end{subequations}
where the explicit expressions of $\omega_{\mathcal{A}}
(\bolds{\theta}, \mu_j )$, $\omega_{\mathcal{B} -
\mathcal{A}}
(\bolds{\theta}, \mu_j )$ and $\omega_{\mathcal{C} -
\mathcal{B}}
(\bolds{\theta}, \mu_j )$ are given in the supplementary material [\citet{SunKuczekZhu2014s1}],
Section~\textup{S.3}.
\end{longlist}
\end{proposition}

From Proposition \ref{teoproposition2}, as $n$ goes to $\infty$, the
calibrated estimates
$\hat{\mu}_j^{xyz}$, $\hat{\mu}_{j}^{yz}$ and $\hat{\mu}_{j}^{z}$
are asymptotically unbiased
estimates of $\mu_j$ for $j$ in $\mathcal{A}$,
$\mathcal{B} - \mathcal{A}$ and $\mathcal{C}-\mathcal{B}$, respectively.
The variances of the calibrated estimates do not
converge to zero asymptotically, instead, they converge to $\gamma
_{\mathcal{A}}$,
$\gamma_{\mathcal{B - A}}$ and $\gamma_{\mathcal{C - B}}$, respectively.
When $n$ is sufficiently large, the second terms in (\ref
{eqscm-prediction-xyz-var}),
(\ref{eqscm-prediction-yz-var}) and (\ref{eqscm-prediction-z-var})
are negligible, and
the variances of the calibrated estimates of $\mu_j$ for $j$ in
$\mathcal{A}$, $\mathcal{B - A}$
and $\mathcal{C - B}$ can be approximated by $\gamma_{\mathcal{A}}$,
$\gamma_{\mathcal{B - A}}$ and
$\gamma_{\mathcal{C - B}}$, respectively. When $n$ is small or
moderate, the second terms in the
above expansions may not be negligible.

It is clear that $\gamma_{\mathcal{B}-\mathcal{A}}$ is larger than
$\gamma_{\mathcal{A}}$, and
$\gamma_{\mathcal{C}-\mathcal{B}}$ is larger than $\gamma_{\mathcal
{B}-\mathcal{A}}$ and
$\gamma_{\mathcal{A}}$. The order implies that when $n$ is
sufficiently large, the calibrated
estimate of $\mu_{j }$ based on only
the measurements of RNA-Seq is less accurate than the calibrated
estimate of $\mu_{j}$ based on
the RNA-Seq and microarray measurements, which, further, is less
accurate than those calibrated
estimates based on all three platforms.

We further compare the three types of calibrated estimates with the
qRT-PCR measurement $X_{j}$, microarray measurement $Y_{j}$ and the
RNA-Seq measurement $Z_{j}$.
Note that ${X}_{j}$, which is presumably the most accurate measurement
among $X_j$, $Y_j$ and
$Z_j$, is an unbiased estimate of $\mu_j$, and its reproducibility is
$\sigma^2_1$. Because
$\gamma_{\mathcal{A}}$ is less than
$\sigma_1^2$, $\hat{\mu}_{j}^{xyz}$ is more accurate than ${X}_{j}$
when $n$ is sufficiently large.
Therefore, by combining the measurements from all three platforms, we
can obtain a more accurate
estimate of the expression level of gene $j$.
Because $\gamma_{\mathcal{B - A}}$ is less than either of
$\sigma_2^2/\beta_2^2$ and $\sigma_3^2/\beta_3^2$, which are the
reproducibilities of $Y_{j}$ and
$Z_{j}$, respectively, $\hat{\mu}_j^{yz}$ is more accurate than
$Y_{j}$ and $Z_{j}$ when $n$ is
large.
$\gamma_{\mathcal{C - B}}$ is equal to $\sigma_3^2/ \beta_3^2$,
which is the reproducibility
of $Z_{j}$. Therefore, $\hat\mu_{j}^{z}$ has the same reproducibility
as the RNA-Seq measurement
$Z_{j}$. The advantage for using $\hat\mu_{j}^{z}$ instead of $Z_j$
is that $\hat\mu_{j}^{z}$ is in
the same scale as $\hat\mu_j^{xyz}$ and $\hat\mu_j^{yz}$ so that the
calibrated estimates in $\mathcal{C - B}$ are
scale compatible with those in
$\mathcal{A}$ and $\mathcal{B - A}$.

The calibrated estimates $\{ \hat\mu^{xyz}_j\dvtx  j \in\mathcal{A}\}$,
$\{\hat\mu^{yz}_j\dvtx  j \in\mathcal{B} - \mathcal{A}\}$ and $\{\hat\mu
^{z}_j\dvtx  j \in\mathcal{C} -
\mathcal{B} \}$ can also be considered gene expression measurements
normalized by the qRT-PCR
platform. As discussed
above, the standard errors of these three types of calibrated estimates
are different.\vspace*{2pt}
Let ${\varphi}_{j, \mathcal{A}}$, ${\varphi}_{j, \mathcal{B-A}}$
and ${\varphi}_{j,
\mathcal{C-B}}$ denote the first order expansions of the variances of
$\hat\mu_{j}^{xyz}$,
$\hat\mu_{j}^{yz}$ and $\hat\mu_{j}^{z}$ in
(\ref{eqscm-prediction-xyz-var}),
(\ref{eqscm-prediction-yz-var}) and (\ref{eqscm-prediction-z-var}),
respectively. By replacing
the structural parameters with their corresponding
estimates, we obtain the estimated variances and standard
errors of the three types of calibrated estimates, which\vspace*{-2pt} are
$\hat{\varphi}_{j, \mathcal{A}}$ and $\sqrt{\hat{\varphi}_{j,
\mathcal{A}}}$,
$\hat{\varphi}_{j, \mathcal{B-A}}$ and $\sqrt{\hat{\varphi}_{j,
\mathcal{B-A}}}$,
and $\hat{\varphi}_{j, \mathcal{C-B}}$ and $\sqrt{\hat{\varphi}_{j,
\mathcal{C - B}}}$,
respectively.
When used for gene expression
analysis such as detecting differentially expressed genes, both the
calibrated estimates and
their standard errors need to be used.

%
\begin{table}[b]
\tabcolsep=10pt
\caption{The settings of model parameters in simulation study}\label{Tabsimulation-setting-single}
\begin{tabular*}{\tablewidth}{@{\extracolsep{\fill}}@{}ld{2.2}d{1.1}d{1.2}d{1.2}d{1.1}d{1.1}d{1.2}@{}}
\hline
& \multicolumn{2}{c}{\textbf{Intercept}} & \multicolumn{2}{c}{\textbf{Slope}} & \multicolumn{3}{c@{}}{\textbf{Reproducibility}}\\[-6pt]
& \multicolumn{2}{c}{\hrulefill} & \multicolumn{2}{c}{\hrulefill} & \multicolumn{3}{c@{}}{\hrulefill}\\
& \multicolumn{1}{c}{$\bolds{\alpha_2}$} & \multicolumn{1}{c}{$\bolds{\alpha_3}$}
& \multicolumn{1}{c}{$\bolds{\beta_2}$} & \multicolumn{1}{c}{$\bolds{\beta_3}$}
& \multicolumn{1}{c}{$\bolds{\sigma_1^2}$} & \multicolumn{1}{c}{$\bolds{\sigma_2^2}$}& \multicolumn{1}{c@{}}{$\bolds{\sigma_3^2}$} \\
\hline
Setting 1 &  9  &  5  &  0.75  &  1  &  0.8  &  1.2  &  1  \\
Setting 2 &  0.02  &  0.2  &  0.9  &  0.95  &  0.5  &  1  &  0.75 \\
Setting 3 &  -5  &  5  &  1.3  &  1.2  &  0.2  &  1  &  1.2  \\
\hline
\end{tabular*}
\end{table}

\section{Simulation}\label{secsimulation}
An extensive simulation study was conducted to evaluate the performance
of the proposed
calibration method. In this section we present the simulation results
under the single-lab
scenario. We first report the simulation results on the accuracy of the
calibrated estimates
or calibrated expression levels, and then we present the performance of
the calibrated
estimates
when used in gene differential expression (DE) analysis.

\subsection{Accuracy of estimates}
In this section we report the simulation results for three settings of
the system of ME models
under the single-lab scenario. The structural parameters of the three
settings are listed in Table~\ref{Tabsimulation-setting-single}. Setting 1 is set to resemble the
results from real data
analysis in Section~\ref{secResults}. In setting 2 the
reproducibilities of microarray and RNA-Seq are mildly worse than that
of qRT-PCR technology; and
in setting 3 the reproducibilities of microarray and RNA-Seq are much
worse that that of qRT-PCR.
The incidental parameters (i.e., $\mu_j$'s) for each model setting
were randomly drawn from $N
(0, 25 )$.
For each model setting, we independently generated a training data set
of 300 genes and a testing
data set of 1000 genes. Both data sets
contain the measurements of the genes by all three platforms.
The measurements of the first $n$ genes in the training set were used
to estimate the structural
parameters ($n$ was varied from $20$ to $300$), and then the three
types of calibrated estimates of
the expression levels
of the genes in the testing set were obtained. For each combination of model
setting and $n$, this procedure was repeated 200 times to calculate the
MSEs of the estimates for
the purpose of performance evaluation and comparison.

We first examine and compare the variances of the calibrated expression
levels $\hat{\mu}_j^{xyz}$,
$\hat{\mu}_j^{yz}$ and $\hat{\mu}_j^{z}$. Recall
that $\operatorname{Var}(\hat{\mu}_j^{xyz})= \gamma_{\mathcal{A}}+n^{-1}
\omega_{\mathcal{A}}(\bolds{\theta}, \mu_j)$. When $n$ is large,
$\gamma_{\mathcal{A}}$
is the dominant term and the other higher order terms are negligible;
but when $n$ is moderate
(e.g., $n \in[20, 50 ]$), the second term $n^{-1} \omega
_{\mathcal{A}}(\bolds{\theta}, \mu_j)$ is
not negligible. Further notice
that $\gamma_{\mathcal{A}}$ depends on the structural parameters
only, but $n^{-1}
\omega_{\mathcal{A}}(\bolds{\theta}, \mu_j)$ also depends on the
incidental parameters, particularly
$(\mu_j-\bar{\mu})^2$ as shown in the supplementary material
[\citet{SunKuczekZhu2014s1}], Section~\textup{S.3}.
This implies that the
variance of $\hat{\mu}_j^{xyz}$ increases as $\mu_j$ deviates away
from the mean $\bar{\mu}$ when
$n$ is moderate. The variances of $\hat{\mu}_j^{yz}$ and $\hat{\mu
}_j^{z}$ behave in the same way as
the variance of $\hat{\mu}_j^{xyz}$.

%
\begin{figure}

\includegraphics{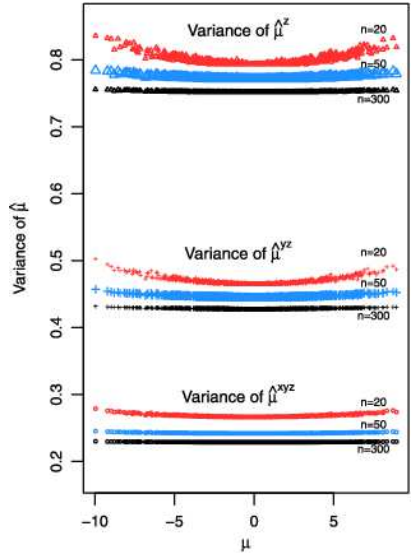}

\caption{Variances of calibrated estimates of gene expression levels.
The bottom three curves (plotted in circles) are for
$ \{\operatorname{Var}(\hat{\mu}_{j}^{xyz}) \}$, the middle three
curves (plotted in crosses) are for
$ \{\operatorname{Var}(\hat{\mu}_{j}^{yz}) \}$ and the top three
curves (plotted in triangles) are for $ \{\operatorname{Var}(\hat{\mu}_{j}^{z})
\}$, with
$n=20, 50, 300$, respectively.}\label{figvar-pred-single}
\end{figure}

Figure~\ref{figvar-pred-single} from bottom to top shows the scatter
plots of the sample variances
of $\hat{\mu}_j^{xyz}$,
$\hat{\mu}_j^{yz}$ and $\hat{\mu}_j^{z}$ versus the incidental parameters
$\mu_j$ for the first model setting and three different sizes of
training data subsets (black
for $n=20$, blue for $n=50$, and red for $300$).
For convenience, we refer to the curve generated by plotting the
variance of a type of estimate
against the incidental parameter as the \textit{variance curve} of the
type of estimate.
The bottom three curves are the variance curves of $\hat{\mu}_j^{xyz}$
for $n=20, 50$ and
$300$, respectively,
the middle three curves are those of $\hat{\mu}_j^{yz}$, and the top
three curves are those of
$\hat{\mu}_j^{z}$.

For each type of estimate, the variance curve for $n=20$ is above that
for $n=50$, which
is above that for $n=300$, indicating that as $n$ increases, the
variance of the respective estimate
decreases. Furthermore, all three variance curves for \mbox{$n=20$}
demonstrate a stronger
quadratic pattern than those for $n=50$ and $n=300$, indicating their
dependence on $(\mu_j-\bar\mu)^2$. The variance curves for $n=50$
show a
much mitigated quadratic pattern, and the variance curves for $n=300$
become flat.
Comparing\vspace*{1pt} the variance curves, it is clear that
under the same $n$, the variance curve of $\hat{\mu}_j^{xyz}$ is
lower than
that of $\hat{\mu}_j^{yz}$, which is lower than that of $\hat{\mu
}_j^{z}$, indicating that, in
terms of accuracy, the order of the three types of estimates, from the
best to the worst, is
$\hat{\mu}_j^{xyz}$, $\hat{\mu}_j^{yz}$ and $\hat{\mu}_j^{z}$.
These results confirm the properties of the variances of the three
types of estimates discussed
in Section~\ref{secModel}.

%
\begin{figure}

\includegraphics[scale=0.99]{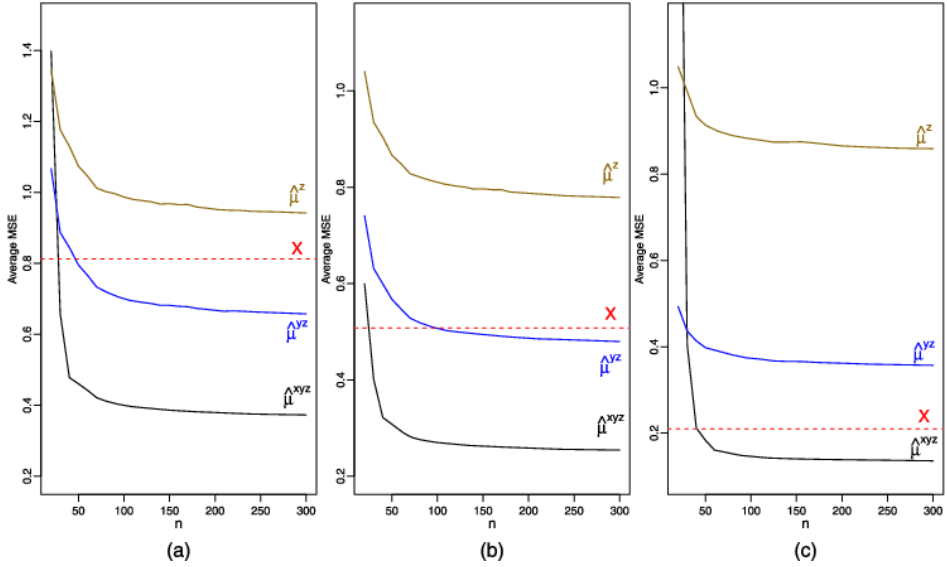}

\caption{Average MSEs of original measurements and calibrated estimates.
Plots \textup{(a)}, \textup{(b)} and \textup{(c)} are for model settings 1, 2 and 3,
respectively.}\label{figmse-mu-single}\vspace*{-3pt}
\end{figure}

We further examine the average MSEs of the three types of calibrated
estimates over all of the genes
in the testing data set.
For convenience in discussion, we refer to the scatter plot of the
average MSE of a type of estimate
over all of the genes in the testing data set versus the training data
size $n$ as the \textit{aMSE
curve} of the type of estimate.
Note that $X_j$ is also an unbiased estimate of $\mu_j$, and we also
compare the average MSEs of $X_j$
with the three types of estimates.
Figure~\ref{figmse-mu-single} demonstrates the aMSE curves of the
three types of estimates
$\{\hat{\mu}_j^{xyz}\}$, $\{\hat{\mu}_j^{yz}\}$ and $\{\hat{\mu}_j^{z}\}
$ as\vspace*{1pt} well as that of the
original
qRT-PCR measurements $\{{X}_{j}\}$.
Plots (a), (b) and (c) in Figure~\ref{figmse-mu-single} from the left
panel to the right panel
correspond to the three model settings given in Table~\ref{Tabsimulation-setting-single},
respectively.

The aMSE curves of $\{{X}_{j}\}$ are flat because they do not
depend on $n$. Overall, the aMSE curves of $\{\hat{\mu}_j^{xyz}\}$,
$\{\hat{\mu}_j^{yz}\}$
and $\{\hat{\mu}_j^{z}\}$ decrease as $n$ increases, and they become
flat after $n$ is greater than
$100$, indicating that the structural parameters are accurately
estimated and the benefit of
further increasing $n$ in the training data set becomes negligible.

In all of the plots, the aMSE curves of $\{\hat{\mu}_j^{xyz}\}$ are
always the lowest, indicating
that $\{\hat{\mu}_j^{xyz}\}$ give the most accurate quantification of
the gene expression levels.
The aMSE curves of $\{\hat{\mu}_j^{yz}\}$ are always below the aMSE
curves of
$\{\hat{\mu}_j^{z}\}$, indicating that $\{\hat{\mu}_j^{yz}\}$ is
more accurate than
$\{\hat{\mu}_j^{z}\}$. However, the aMSE curve of $\{\hat{\mu}_j^{yz}\}
$ is
not\vspace*{1.5pt} always below that of $\{{X}_{j}\}$. Plots (a), (b) and (c)
demonstrate three
different cases where when $n$ is sufficiently large (e.g., $\ge$100),
the aMSE curves of
$\{\hat{\mu}_{j}^{yz}\}$ are below, close to and above that of $\{
{X}_{j}\}$. In general, the
performance of $\{\hat{\mu}_{j}^{yz}\}$ relative to $\{{X}_{j}\}$
depends on the
model settings.
In all of the plots, the aMSE curves of $\{\hat{\mu}_j^{z}\}$ are
always above that
of $\{{X}_{j}\}$, indicating that the calibrated estimate based on
RNA-Seq measurement alone cannot
outperform the qRT-PCR measurement.\looseness=1

As discussed in Section~\ref{secModel}, the structural parameters are
fundamentally different from
the incidental parameters. As $n$ increases, the estimates of the
structural parameters will
converge to their respective targets as shown in Proposition \ref
{teoproposition1}. The simulation
results about the
structural parameters can be found in the supplementary material
[\citet{SunKuczekZhu2014s1}], Section~\textup{S.4}.

\subsection{Differential expression}\label{secsimulationde}
In this section we report the simulation study on the performance of
calibrated estimates when
used in DE
analysis. In order to make the simulation study convincing, we
simulated the RNA-Seq data from a
popularly used simulator called the Flux Simulator [\citet
{Griebel2012}]. The Flux Simulator mimics
the pipeline of RNA-Seq experiments and generates RNA-Seq short reads.
It allows researchers to
investigate the properties of RNA-Seq data and the analysis tools \textit{in silico}.
In this study, we considered two biological conditions, which are
referred to as conditions 1~and~2. Totally, $5000$ genes were considered, and among them
$500$ genes were set to be
differentially expressed. Let $\mu_{1j}$ and $\mu_{2j}$ denote the
log-2 values of the true
expression levels of gene $j$ under the two conditions, respectively,
and $d_j = \mu_{j1} -
\mu_{j2}$ the difference between $\mu_{1j}$ and $\mu_{2j}$.
Under condition 1, $\mu_{1j}$'s were randomly assigned by the Flux
Simulator using its default
setting. When gene $j$ is not differentially expressed between the two
conditions, we set $d_j$ to
be $0$; otherwise, $d_j$ is randomly generated from the uniform
distribution on $[-2, -0.5] \cup
[0.5, 2]$. Then under condition 2, $\mu_{2j}$'s were set to be $\mu
_{1j} + d_j$.
For each biological sample, 5 million reads were generated and mapped
back to the
reference genome. Finally, the log-2 RPKM values were calculated and
used as the RNA-Seq
expression measurements. qRT-PCR and microarray data were generated
from the
system of ME models, with the true expression levels on the two
conditions set the same as those
used for the RNA-Seq data. The structural parameters related to qRT-PCR
and microarray (i.e.,
$\alpha_2$, $\beta_2$, $\sigma_1$ and $\sigma_2$) were set the
same as setting 1 in Table~1. We
generated qRT-PCR data for $500$ genes and microarray data for $3000$
genes. Therefore, in this
simulation study, there are $500$ genes in set $\mathcal{A}$, $3000$
genes in set $\mathcal{B}$,
and $5000$ gene in set $\mathcal{C}$.

After generating the expression measurements, we applied the system of
ME models for each
biological condition. The calibrated expression levels and their
variances are\vspace*{2pt} $\hat\mu_{j}^{xyz}$
and
$\hat\varphi_{j, \mathcal{A}}$, $\hat\mu_{j}^{yz}$ and $\hat
\varphi_{j, \mathcal{B-A}}$, and
$\hat\mu_{j}^{z}$ and $\hat\varphi_{j, \mathcal{C-B}}$, for $j \in
\mathcal{A}$, $\mathcal{B-A}$ and
$\mathcal{C-B}$, respectively. In what follows, for convenience, we do
not further distinguish the
notation for these three types of calibrated estimates. Instead, we use
$\hat\mu_{j1}$ and
$\hat\mu_{j2}$ to denote the calibrated expression levels of gene $j$
under condition 1 and
condition 2, respectively, and use $\hat\varphi_{j1}$ and $\hat
\varphi_{j2}$ to denote the
estimated variances of $\hat\mu_{j1}$ and $\hat\mu_{j2}$,
respectively.\looseness=1

We performed DE analysis for the two conditions using the $z$-test
based on the
RNA-Seq measurements ($Z_j$'s) and the calibrated measurements ($\hat\mu
_j$'s) separately, and
compared the results. When performing the $z$-test based on the
calibrated measurements for gene $j$,
the p-value
for testing $H_0\dvtx  \mu_{j1} = \mu_{j2}$ was
\begin{equation}
\label{eqde-pred-pvalue} p_j = 2 P \biggl(Z > \frac{\llvert\hat{\mu
}_{j1} - \hat{\mu}_{j2}\rrvert}{\sqrt{\hat\varphi_{j1} + \hat\varphi
_{j2}}} \biggr),
\end{equation}
where $Z$ follows the standard normal distribution. Based on the
calculated \mbox{$p$-}values, the standard
Benjamini--Hochberg procedure [\citet{BenjaminiHochberg1995}] was used
to identify differentially
expressed genes at a given false discovery rate (FDR). The $z$-test
based on the RNA-Seq
measurements was similarly conducted.

%
\begin{figure}

\includegraphics{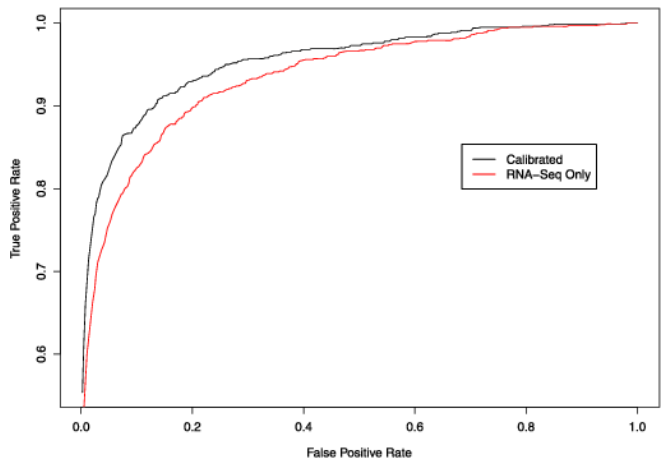}

\caption{ROC curves of DE analysis based on the calibrated
expression levels and the RNA-Seq measurements.}\label{figsimulation-roc}
\end{figure}

Figure~\ref{figsimulation-roc} depicts the Receiver Operator
Characteristic (ROC) curves comparing
the performances of the calibrated expression measurements and the
RNA-Seq expression measurements
in DE analysis.
The ROC curve of the calibrated expression levels (black line) is above
that of the RNA-Seq
expression measurements (red line), indicating the DE results based on the
calibrated expression measurements have higher true positive rate than
those based on the
uncalibrated RNA-Seq expression measurements at any level of false
positive rate. For example,
at the false positive rate level of $0.05$, the true positive rate of
the uncalibrated
RNA-Seq measurements was $0.565$, while
that of the calibrated expression measurements was $0.641$.
Therefore, the calibrated expression measurements outperform the
uncalibrated RNA-Seq
measurements in this example.

\section{Application and results}\label{secResults}

\subsection{Data sets}
\label{secDatasets}

We applied the system of ME models to analyze the qRT-PCR,
RNA-Seq and microarray gene expression data of two RNA samples, namely,
the human brain reference RNA sample (or, in short,
the Brain sample) and the human universal reference RNA sample (or the
UHR sample), generated by
the Microarray Quality Control (MAQC) and Sequencing Quality Control
(SEQC) projects.\vspace*{1pt}

The qRT-PCR data were generated by one single lab using
TaqMan\tsup{\textregistered} Gene Expression
Assays and can be downloaded from Gene Expression Omnibus (GEO) with
the series number GSE5350
(\url{ftp://ftp.ncbi.nih.gov/pub/geo/DATA/supplementary/series/GSE5350/}).
The qRT-PCR data\break  were
normalized by the delta--delta Ct method and originally contain the
qRT-PCR measurements of $1001$
genes; see the supplementary material Section~S1.2 of
\citet{Bullard2010} for detailed
information. Among the $1001$ genes, $7$ genes have multiple entries with
distinct expression values but under the same RefSeq ID. To avoid
ambiguity, these genes were
removed. Each gene has $4$ technical replicates for each of
the UHR and Brain samples.

The RNA-Seq data include measurements generated from two RNA-Seq
experiments conducted in two
different labs
using the Illumina Genome Analyzer. The data from the first lab
[\citet{Bullard2010}] can be downloaded from the NCBI Sequence
Read Archive (SRA)
(\url{http://www.ncbi.nlm.nih.gov/sra/}) under the accession number
SRA010153, and the data from the
second lab can be downloaded from the same website under the accession
number SRA008403. For
convenience, we will use the accession numbers to refer to these two
data in the rest of the
article. The RPKM value is calculated lane-by-lane for each gene. For
each of the Brain and UHR
samples, those genes that received no reads in at least one lane of the
two RNA-Seq experiments
were excluded.
See the supplementary material [\citet{SunKuczekZhu2014s1}],
Section~\textup{S.7}, for more information about
the two RNA-Seq data
sets.

The microarray data were generated by five different microarray
experiments conducted in
five different labs using Affymetrix U133 Plus2.0. For convenience, we
label the labs as MA1 to MA5
in the rest of the article.
The original probe-level data can be downloaded from Gene Expression
Omnibus (GEO) with the series
number GSE5350 (\url
{ftp://ftp.ncbi.nih.gov/pub/geo/DATA/supplementary/series/GSE5350/}).
Data from
each lab have 5 replicates for both UHR and Brain samples, and was
normalized using the PLIER
[\citet{PLIERWhite}] method. Detailed information about these
data sets can be found in the
website of the MAQC projects
(\url{http://www.fda.gov/}). For each replicate in each
lab, the average probe-level measurement of a gene is considered the
gene's expression level
intensity.

We integrated the three different types of gene expression data as
follows. First, the log-2
transformation was applied to all three types of gene expression data.
Then for each gene and each
lab, the mean measurement across technical replicates is used as the
gene's expression value.
Furthermore, we exclude genes with extremely low or extremely high
expression levels (i.e., those with qRT-PCR expression measurements
below $-6$ or above $4$ in log-2
scale) in
$\mathcal{A}$; the remaining genes in $\mathcal{A}$ are used to
estimate the structural parameters.
For the Brain sample, there are $409$ genes with
expression data from all three platforms ($\mathcal{A}$), $6419$ genes
with only RNA-Seq and
microarray
measurements ($\mathcal{B}-\mathcal{A}$), and $5949$ genes with only
RNA-Seq measurements
($\mathcal{C}-\mathcal{B}$); and for the UHR sample,
there are $477$ genes with measurements by all three platforms
($\mathcal{A}$),
$7187$ genes with measurements only by RNA-Seq and microarray
($\mathcal{B}-\mathcal{A}$),
and $6892$ genes with only measurements by RNA-Seq ($\mathcal
{C}-\mathcal{B}$).

Two schemes were used to analyze the integrated data. First, we
considered the single-lab scenario,
in which each platform has data from one lab. In total, there are ten
possible combinations, and we
applied the system of ME models (\ref{eqmodel-single-pcr})--(\ref
{eqmodel-single-rnaseq}) to each
combination. The analysis results of
individual combinations were similar, and we report only the results
for the combination
that includes the RNA-Seq data from SRA010153 and the microarray data
from MA1.
Second, we applied the system of ME models for the multi-lab scenario
to analyze
the entire data set for each RNA sample. Due to limited space, the
results from the multi-lab
scenario are presented and briefly discussed in the supplementary material
[\citet{SunKuczekZhu2014s1}], Section~\textup{S.8}.

\subsection{Diagnostics of model assumptions}
The system of ME models\break  \mbox{(\ref{eqmodel-single-pcr})--(\ref
{eqmodel-single-rnaseq})} imposes
the normality
and homoscedasticity assumptions on the measurement errors.
We checked these assumptions using genes
in $\mathcal{A}$. After the system of ME models was fitted, we
calculated the residuals
by $e_{1j} = X_{j} - \hat\mu_{j}$, $e_{2j} = Y_{j} - \hat\alpha_2
- \hat{\beta}_2 \hat\mu_{j}$ and
$e_{3j} = Z_{j} - \hat\alpha_3 - \hat{\beta}_3 \hat\mu_{j}$
corresponding to the\vadjust{\goodbreak} measurement errors
due to the qRT-PCR, microarray and RNA-Seq platforms, respectively. To
check the normality
assumption, we
generated the $QQ$ plots
for the residuals and did not detect significant violation of the assumption.
To check the homoscedasticity assumption, we generated residual plots
and constructed approximate
95\% confidence intervals of the Box--Cox transformation. Because the
diagnostic results are similar
for the two RNA samples, we present only those from the UHR sample as
an example. The residual
plots corresponding to the qRT- PCR, microarray and RNA-Seq platforms
are presented in Figure S3 in
the supplementary material [\citet{SunKuczekZhu2014s1}], Section~\textup{S.8}. The plots do not demonstrate
strong heteroscedastic
patterns. The 95\% confidence intervals of Box--Cox transformation for
the residuals from the three
platforms are
presented in Figure S4 in the supplementary material [\citet
{SunKuczekZhu2014s1}], Section~\textup{S.8}. The
confidence intervals
corresponding to the qRT-PCR and RNA-Seq platforms both contain
$1$, and $1$ is on the boundary of the confidence interval
corresponding to the
microarray platform. These results together indicate that there does
not exist significant violation of the homoscedasticity assumption
imposed on the platform
measurement errors.

\subsection{Structural parameters}
The estimates of the structural parameters for the Brain and UHR
samples are given in
Table~\ref{Tabmodel-single-esti}. The standard errors of the estimates
are also reported in
parentheses in the table.
The estimates of the structural parameters can be used to compare the
three platforms in terms of
the quality of measurements they provide.

%
\begin{table}[b]
\tabcolsep=0pt
\caption{Estimates of structural parameters using the qRT-PCR data, RNA-Seq data from SRA010153 and microarray data from MA1}\label{Tabmodel-single-esti}
\begin{tabular*}{\tablewidth}{@{\extracolsep{\fill}}@{}lccccccc@{}}
\hline
& \multicolumn{2}{c}{\textbf{Intercept}} & \multicolumn{2}{c}{\textbf{Slope}} & \multicolumn{3}{c@{}}{\textbf{Variance}} \\[-6pt]
& \multicolumn{2}{c}{\hrulefill} & \multicolumn{2}{c}{\hrulefill} & \multicolumn{3}{c@{}}{\hrulefill}\\
& \multicolumn{1}{c}{$\bolds{\hat{\alpha}_2}$} & \multicolumn{1}{c}{$\bolds{\hat{\alpha}_3}$}
& \multicolumn{1}{c}{$\bolds{\hat{\beta}_2}$} & \multicolumn{1}{c}{$\bolds{\hat{\beta}_3}$}
& \multicolumn{1}{c}{$\bolds{\hat{\sigma}_1^2}$} & \multicolumn{1}{c}{$\bolds{\hat{\sigma}_2^2}$} & \multicolumn{1}{c@{}}{$\bolds{\hat{\sigma}_3^2}$}\\
\hline
Brain &  8.8401  &  4.9405  &  0.7754  &  1.0254  &  0.7945  &  1.2407  &  1.0679  \\
&  (0.0921)  &   (0.1115)  &  (0.0411)  &  (0.0514)  &  (0.1156)  & (0.0955)  &  (0.1641) \\[3pt]
UHR &  9.1033  &  5.4249  &  0.7695  &  1.0009  &  0.6685 &  1.0444  &  0.9452  \\
&  (0.0745)  &  (0.0875)  &  (0.0292) &  (0.0347)  &  (0.0882)  &  (0.0768)  &  (0.1233) \\
\hline
\end{tabular*}
\end{table}

From models (\ref{eqmodel-single-pcr})--(\ref{eqmodel-single-rnaseq}),
the qRT-PCR measurements
are unbiased with respect to $\mu_j$'s, and the intercepts $\alpha_2$
and $\alpha_3$ represent the
shifts of microarray and RNA-Seq measurements relative to
the qRT-PCR measurements. Larger absolute values of $\alpha_2$ and
$\alpha_3$ indicate larger
shifts. From Table~2, $\hat\alpha_2$ are $8.8401$ and $9.1033$ in the
Brain and UHR samples,
respectively; and $\hat\alpha_3$ are $4.9405$ and $5.4249$ in the
Brain and UHR samples,
respectively. In both samples, $\hat\alpha_2 > \hat\alpha_3$,
indicating that microarray
measurements have larger shifts than the RNA-Seq measurements.
The slopes $\beta_2$ and $\beta_3$ represent the scales of microarray
and RNA-Seq measurements
relative to qRT-PCR measurements. From Table~\ref{Tabmodel-single-esti}, $\hat{\beta}_2$ are
$0.7754$ and $0.7695$ in the
Brain and UHR samples, respectively, both of which are significantly
less than~$1$, indicating that
the microarray measurements are in a smaller scale compared to the
qRT-PCR measurements. On the
other hand, $\hat{\beta}_3$ are $1.0254$ and $1.0009$ in the Brain and
UHR samples, respectively,
both of which are not significantly different from $1$, indicating that
the RNA-Seq measurements are
in a similar scale as the qRT-PCR measurements.
The variances $\sigma_1^2$, $\sigma_2^2/\beta_2^2$ and
$\sigma_3^2/\beta_3^2$ reflect the reproducibilities of the three
platforms. Smaller value
indicates higher reproducibility. From Table~2, in the Brain sample, the
three values are
$0.7945$, $2.0635$ and $1.0157$; and in the UHR sample, the three
values are $0.6685$, $1.7638$ and
$0.9435$. In both samples, $\hat\sigma_1^2< \hat\sigma_3^2/\hat{\beta
}_3^2 < \hat
\sigma_2^2/\hat{\beta}_2^2$, indicating that qRT-PCR has the best
reproducibility,
microarray has the worst reproducibility, and RNA-Seq is slightly worse
than qRT-PCR but much
better than microarray.\vspace*{-3pt}

\subsection{Incidental parameters}
After the estimates of the structural parameters were obtained, we used
the formulas obtained
in Section~\ref{secModel} to estimate the gene expression levels, and the
standard errors of the calibrated gene expression values were also
calculated. These calibrated gene
expression values together with their standard errors are expected to
have a higher quality than
the original measurements by the three platforms and lead to better
results in downstream gene
expression analysis.

\begin{figure}

\includegraphics{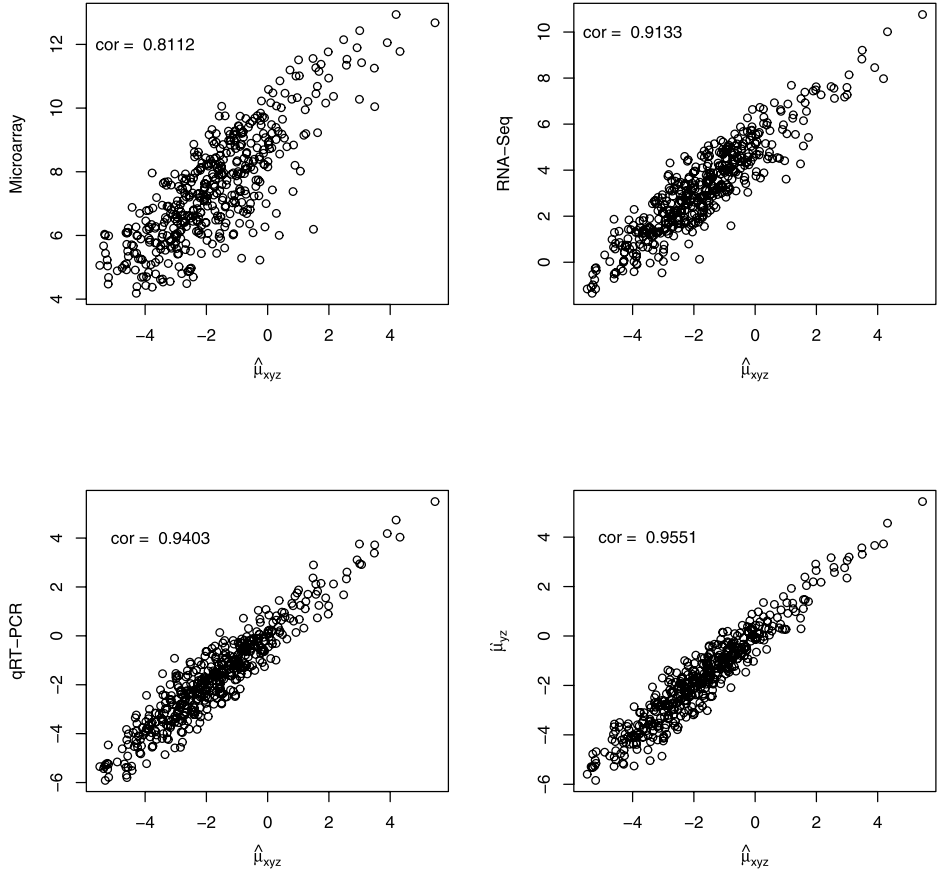}

\caption{Scatter plots of the measurements by microarray, RNA-Seq,
qRT-PCR and
$\hat{\mu}^{yz}$ versus $\hat{\mu}^{xyz}$ on $\mathcal{A}$ in the
Brain sample.}\label{figpredbrain}
\end{figure}

\begin{figure}

\includegraphics{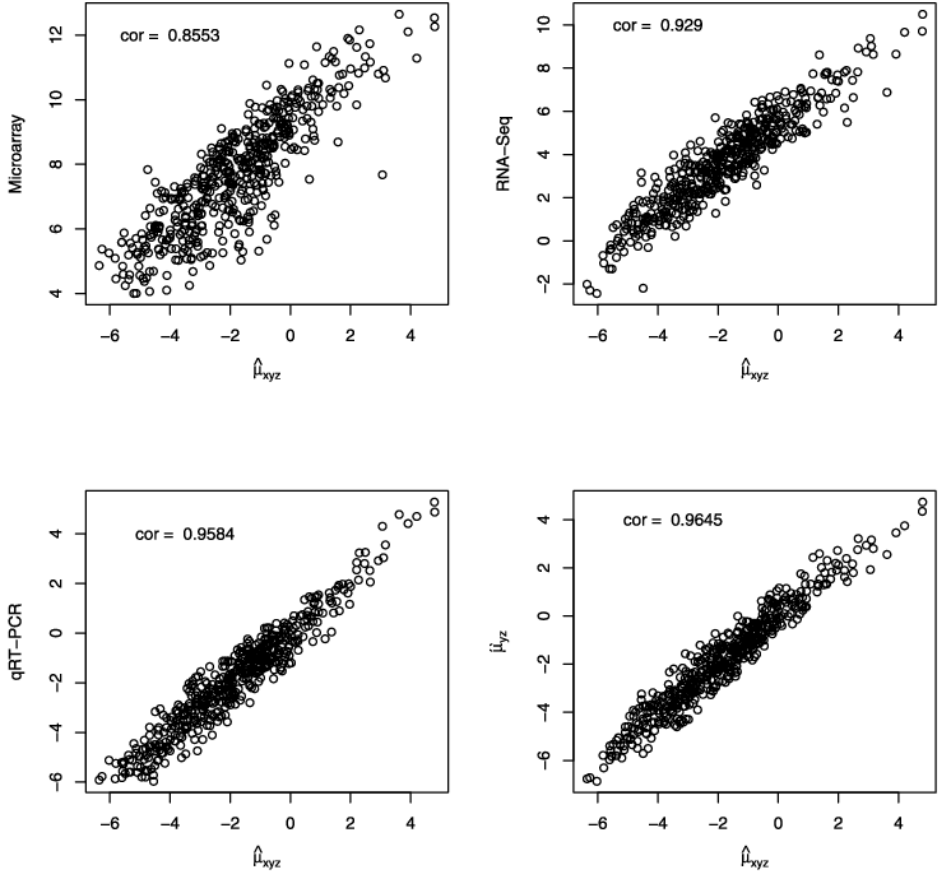}

\caption{Scatter plots of the measurements by microarray, RNA-Seq,
qRT-PCR and
$\hat{\mu}^{yz}$ versus $\hat{\mu}^{xyz}$ on $\mathcal{A}$ in the
UHR sample.}\label{figpreduhr}
\end{figure}

We compare the calibrated gene expression levels with their original
qRT-PCR, microarray and
RNA-Seq measurements, and use the genes in the Brain and UHR samples
that have measurements by all
three platforms as an illustrative example. Because these genes have
measurements by all\vspace*{1pt} three
platforms, all three types of calibrated estimates $\{\hat{\mu
}_{j}^{z}\}$,
$\{\hat{\mu}_j^{yz}\}$ and $\{\hat{\mu}_j^{xyz}\}$ could be calculated.
Based on the theoretical and simulation results in Sections~\ref{secModel} and
\ref{secsimulation}, $\{\hat{\mu}_j^{xyz}\}$ are the\vspace*{1.5pt}
most accurate measurements.
We plotted the original measurements $\{{X}_{j}\}$, $\{{Y}_{j}\}$
and $\{{Z}_{j}\}$ as well as the calibrated measurements $\{\hat{\mu
}_j^{yz}\}$ against
$\{\hat{\mu}_j^{xyz}\}$, and presented the plots in Figures~\ref{figpredbrain}~and~\ref{figpreduhr} for the Brain and UHR sample, respectively.
Because\vspace*{1pt} $\{\hat{\mu}_j^{z}\}$ are the linear transformation of the
original RNA-Seq\vspace*{1pt}
measurements $\{{Z}_{j}\}$, the plot of $\{\hat{\mu}_j^{z}\}$ versus
$\{\hat{\mu}_j^{xyz}\}$
had the same appearance as the plot of $\{{Z}_{j}\}$ versus $\{\hat{\mu
}_j^{xyz}\}$ and thus
was not presented. In each plot, the correlation coefficient between
the two plotted measurements
was calculated and reported.

%

%

From Figure~\ref{figpredbrain}, in the Brain sample, in terms of the
linear correlation
coefficient with the best measurements
$\{\hat{\mu}_{j}^{xyz}\}$, microarray is the worst ($0.8112$) and
qRT-PCR is the best ($0.9403$)
among the three platforms. This is consistent with the findings
reported in the literature that in
terms of the quality of the gene expression data produced by the three
platforms, qRT-PCR is most
accurate and microarray is least accurate. The calibrated\vspace*{-1pt} expression levels
$\{\hat{\mu}_{j}^{yz}\}$ have a higher correlation coefficient with
$\{\hat{\mu}_{j}^{xyz}\}$
($0.9551$) than the qRT-PCR, microarray and RNA-Seq measurements.
In the UHR\vspace*{1pt} sample, in terms of the linear correlation coefficient
with the best measurements $\{\hat{\mu}_{j}^{xyz}\}$, again
microarray is the worst ($0.8553$) and
qRT-PCR is the best ($0.9583$) among the three platforms. The
calibrated expression levels
$\{\hat{\mu}_{j}^{yz}\}$ have a higher correlation coefficient with
$\{\hat{\mu}_{j}^{xyz}\}$
($0.9645$) than qRT-PCR, microarray and RNA-Seq measurements.

\subsection{Gene differential expression}
To demonstrate that calibrated expression
levels can lead to better gene differential expression results, we used
the \mbox{RNA-Seq} measurements
($Z_j$'s) and the calibrated measurements ($\hat\mu_j$'s),
separately, to perform gene
DE analysis for the Brain and UHR samples.

We carried out the DE analysis using the same $z$-test procedure as
described in Section~\ref{secsimulationde}.
Controlling the FDR at 0.01, the numbers of identified genes by the two
types of measurements are
reported in Table~\ref{Tabde-fdr-seperate}.
In total, $331$ genes were detected to be differentially expressed in
the UHR and
Brain samples using the calibrated expression measurements, whereas
only $158$ genes were detected
using the
original RNA-Seq measurements. These two groups of genes share $153$
genes. Therefore, the
calibrated expression measurements led to
the detection of almost all the genes ($153$ out of $158$ genes)
detected by the original RNA-Seq
measurements. In addition, the former detected $173$ more genes than the
latter.
The total number of genes detected by each type of measurements was
further broken down according
to
whether a gene
belongs to $\mathcal{A}$, $\mathcal{B} - \mathcal{A}$ or $\mathcal
{C} - \mathcal{B}$, and the
results are also reported in Table~\ref{Tabde-fdr-seperate}. From the
table, $27$, $132$ and $14$
more genes were detected
by the calibrated expression measurements in $\mathcal{A}$, $\mathcal
{B} - \mathcal{A}$ and
$\mathcal{C - B}$, respectively, than by the original RNA-Seq
measurements. In this case, there
does not exist a gold standard to further verify the selected genes by
the calibrated expression
measurements and the RNA-Seq expression measurements. However, based on
the simulation study in
Section~\ref{secsimulationde}, we believe that the DE analysis based
on the calibrated
measurements has higher true
positive rate and can lead to
more discoveries than the uncalibrated RNA-Seq measurements. A list of
the identified genes by
the calibrated expression measurements is given in the supplementary material
[\citet{SunKuczekZhu2014s1}], Section~\textup{S.9}.

%
\begin{table}
\tabcolsep=0pt
\tablewidth=200pt
\caption{Numbers of differentially expressed genes detected by calibrated estimates and RNA-Seq measurements}\label{Tabde-fdr-seperate}
\begin{tabular*}{\tablewidth}{@{\extracolsep{\fill}}@{}lccc@{}}
\hline
\textbf{Gene set} & \textbf{Calibration} & \textbf{RNA-Seq} & \textbf{Overlap}\\
\hline
$\mathcal{A}$ & \phantom{0}36 & \phantom{00}9 & \phantom{00}9 \\
$\mathcal{B} - \mathcal{A}$ & 192 & \phantom{0}60 & \phantom{0}60 \\
$\mathcal{C} - \mathcal{B}$ & 103 & \phantom{0}89 & \phantom{0}84 \\[3pt]
Total & 331 & 158 & 153 \\
\hline
\end{tabular*}
\end{table}

\section{Discussion}\label{secDiscussion}

A system of functional measurement error (ME) models was proposed to
calibrate the microarray
and RNA-Seq measurements of gene expression levels by qRT-PCR.
Due to limited space, the design issue of the proposed approach was not
discussed in this article.
The success of the proposed approach hinges on the genes that are
measured by all three platforms,
and the major bottleneck is the relative low throughput of the qRT-PCR
platform. Therefore,
the design issue is centered on the qRT-PCR platform with respect to
two questions. The first
question is how many genes
should to be measured by qRT-PCR, and the second question is which
genes should be
measured. For the first question, based on the theoretical, simulation
and real data application
results in this article, it seems that at least 150 genes are needed to
ensure that the bias and
variances (i.e., the
structural parameters) can be accurately estimated and the calibrated
gene expression levels (i.e.,
the incidental
parameters) can reach their best possible accuracy. Vendors of the
qRT-PCR platform such as Life
Technologies now offer assays and services to measure a sufficiently
large number of genes
simultaneously. This makes our proposed approach
feasible in practice. For the second question, based on our theoretical
results (e.g., Proposition~\ref{teoproposition1}), the true expression
levels of the genes selected to be measured by qRT-PCR should be as
spread out as
possible. These two questions and the design issue in general will be
addressed in a future
publication.

As discussed in the \hyperref[secIntroduction]{Introduction}, RNA-Seq data are also found to be
subject to excessive variability
and various methods
have been proposed to normalize \mbox{RNA-Seq} data. In this article, only the
RPKM method and
the resulting RPKM measurements were used as the RNA-Seq measurements.
Clearly, other normalization
methods and their resulting measurements can
also be considered. The variance component due to the RNA-Seq platform
estimated under the multi-lab
scenario (i.e., $\varsigma_2^2$) in this article corresponds to the
RPKM method. If another
normalization method (e.g., the ``mseq'' method) is used, then
$\varsigma_2^2$ will correspond to
that normalization method. Therefore, the system of ME models can be
used to compare different
normalization methods.

The functional system of ME models of order 3 can be extended to that
of order $p > 3$, and the
two-step approach to
parameter estimation can also be extended in a straightforward manner.
In this article, the bias of the measurement of a platform is assumed
be a linear function of the
true expression level. In general, nonlinear models or even
nonparametric models can be considered.
Furthermore,
covariates can also be incorporated into the system of functional ME
models. These possible
extensions of the proposed approach will be investigated in the near future.

%
\section*{Acknowledgments}
We thank the Associate Editor and the reviewers for their constructive
comments and suggestions that helped to improve our manuscript.

\begin{supplement}\label{Supp}
\stitle{Supplement to ``Statistical calibration of \mbox{qRT-PCR}, microarray and RNA-Seq
gene expression data with measurement error models''\\}
\slink[doi]{10.1214/14-AOAS721SUPP} 
\sdatatype{.pdf}
\sfilename{AOAS721\_supp.pdf}
\sdescription{We provide additional supporting materials on the proof
and derivation of Propositions \ref{teoproposition1} and \ref{teoproposition2}, simulation results on the single-lab scenario,
description of the
multi-lab scenario, results from the multi-lab scenario and a list of
differentially expressed
genes by the calibrated measurements.}
\end{supplement}



\printaddresses

\end{document}